\documentclass[aps,pra,twocolumn,superscriptaddress,nofootinbib,amsmath]{revtex4-1}
\usepackage{graphicx}
\usepackage[colorlinks,urlcolor=blue,citecolor=blue,linkcolor=blue]{hyperref}
\usepackage{color}

\usepackage{stackengine}
\stackMath

\begin{document}


\title{Measurement of the dynamic polarizability of Dy atoms near the 626-nm intercombination line}


\author{Marian Kreyer}
\affiliation{Institut f{\"u}r Experimentalphysik and Forschungszentrum f{\"u}r Quantenphysik, Universit{\"a}t Innsbruck, 6020, Innsbruck, Austria}

\author{Jeong Ho Han}
\altaffiliation{Now at Korea Research Institute of Standards and Science, Daejeon 34113, South Korea}
\affiliation{Institut f{\"u}r Experimentalphysik and Forschungszentrum f{\"u}r Quantenphysik, Universit{\"a}t Innsbruck, 6020, Innsbruck, Austria}
\affiliation{Institut f{\"u}r Quantenoptik und Quanteninformation (IQOQI), {\"O}sterreichische Akademie der Wissenschaften, 6020, Innsbruck, Austria}


\author{Cornelis Ravensbergen}
\affiliation{Institut f{\"u}r Experimentalphysik and Forschungszentrum f{\"u}r Quantenphysik, Universit{\"a}t Innsbruck, 6020, Innsbruck, Austria}

\author{Vincent Corre}

\affiliation{Institut f{\"u}r Experimentalphysik and Forschungszentrum f{\"u}r Quantenphysik, Universit{\"a}t Innsbruck, 6020, Innsbruck, Austria}
\affiliation{Institut f{\"u}r Quantenoptik und Quanteninformation (IQOQI), {\"O}sterreichische Akademie der Wissenschaften, 6020, Innsbruck, Austria}

\author{Elisa Soave}
\affiliation{Institut f{\"u}r Experimentalphysik and Forschungszentrum f{\"u}r Quantenphysik, Universit{\"a}t Innsbruck, 6020, Innsbruck, Austria}

\author{Emil Kirilov}
\affiliation{Institut f{\"u}r Experimentalphysik and Forschungszentrum f{\"u}r Quantenphysik, Universit{\"a}t Innsbruck, 6020, Innsbruck, Austria}
\affiliation{Institut f{\"u}r Quantenoptik und Quanteninformation (IQOQI), {\"O}sterreichische Akademie der Wissenschaften, 6020, Innsbruck, Austria}

\author{Rudolf Grimm}
\affiliation{Institut f{\"u}r Experimentalphysik and Forschungszentrum f{\"u}r Quantenphysik, Universit{\"a}t Innsbruck, 6020, Innsbruck, Austria}
\affiliation{Institut f{\"u}r Quantenoptik und Quanteninformation (IQOQI), {\"O}sterreichische Akademie der Wissenschaften, 6020, Innsbruck, Austria}


\date{\today}

\begin{abstract}
We report on measurements of the anisotropic dynamical polarizability of Dy near the 626-nm intercombination line, employing modulation spectroscopy in a one-dimensional optical lattice. To eliminate large systematic uncertainties resulting from the limited knowledge of the spatial intensity distribution, we use K as a reference species with accurately known polarizability. This method can be applied independently of the sign of the polarizability, i.e., for both attractive and repulsive optical fields on both sides of a resonance. By variation of the laser polarization we extract the scalar and the tensorial part. To characterize the strength of the transition, we also derive the natural linewidth. We find our result to be in excellent agreement with literature values, which provide a sensitive benchmark for the accuracy of our method. In addition we demonstrate optical dipole trapping on the intercombination line, confirming the expected long lifetimes and low heating rates. This provides an additional tool to tailor optical potentials for Dy atoms and for the species-specific manipulation of atoms in the Dy-K mixture.
\end{abstract}


\maketitle

\section{Introduction}

Ultracold gases of submerged-shell lanthanide atoms (Dy, Ho, Er, Tm) have emerged as novel platforms for exploring the exciting many-body physics of exotic states of quantum matter under well defined and widely controllable conditions. The intriguing properties of such strongly magnetic atoms result from long-range anisotropic interactions in combination with tunability of the contact interaction. Prominent examples for novel states of matter created in the laboratory are quantum ferrofluids of Dy \cite{Kadau2016otr} and supersolids realized with both Dy and Er \cite{Tanzi2019ooa, Bottcher2019tsp, Chomaz2019lla}.
Progress has also been made with quantum-gas mixtures of different lanthanide atoms (Dy-Er) \cite{Trautmann2018dqm, Durastante2020fri} and mixtures of lanthanide and alkali-metal atoms (Dy-K) \cite{Ravensbergen2018poa, Ravensbergen2020rif}, representing intriguing systems that offer wide potential for future applications.

Submerged-shell lanthanide atoms offer a multitude of optical transitions, which provide flexible tools for efficient laser cooling and trapping \cite{Lu2010tud, Lu2011soa, Frisch2012nlm, Maier2014nlm} and which open up a broad range of applications based on the optical manipulation of atoms. Examples include optical pumping \cite{Schmitt2013soa}, the excitation of Rydberg states \cite{Hostetter2015moh}, realization of spin-orbit coupling \cite{Burdick2016lls}, atomic clock applications \cite{Golovizin2019isc}, quantum-enhanced sensing \cite{Chalopin2018qes, Evrard2019ems}, and quantum spin models \cite{Makhalov2019pqc}. The wide range of applications has motivated theoretical \cite{Dzuba2011dpa, Lepers2014aot, Li2017oto, Li2017aot} and experimental \cite{Lu2011sdb, Maier2015PhD, Schmitt2017PhD, Sukachev2016ism, Golovizin2017mfd, Becher2018apo, Ravensbergen2018ado,Chalopin2018als} studies on the dynamic polarizability, which is the key quantity that characterizes the strength of the atomic interaction with laser light. Because of the complicated electronic structure accurate theoretical models are very challenging and can be refined based on experimental data.

In our recent work \cite{Ravensbergen2018ado}, we introduced a method that greatly improves the accuracy of measurements of the real part of the ground-state dynamic polarizability based on optical dipole potentials \cite{Grimm2000odt}. The basic principle is a comparison of the optical response of the species under investigation with the response of a reference species to the same light field~\cite{Neyenhuis2012apo,Danzl2010auh}. As the key point, this method eliminates uncertainties caused by the limited knowledge of spatial light intensity distribution.
In Ref.~\cite{Ravensbergen2018ado} we demonstrated a polarizability measurement for Dy atoms with K atoms as a reference species by observing collective oscillations in near-infrared light. However, such a collective-excitation scheme can be applied only if the dynamical polarizabilities of both species are positive, i.e., if the laser light attracts and traps the atoms. This limitation substantially reduces the optical wavelength range where the method can be applied. 

In this article, we introduce a more general scheme to measure the dynamical polarizability, which relies on the same basic principle as introduced in Ref.~\cite{Ravensbergen2018ado} but is independent of the sign of the polarizability. Instead of observing collective oscillations of trapped atoms, we use modulation spectroscopy in an optical lattice~\cite{Denschlag2002abe, Heinze2011mso}, applicable for both attractive and repulsive light. As a case study, we investigate the dynamical polarizability near the 626-nm intercombination line of Dy, which is widely used for narrow-line laser cooling \cite{Maier2014nlm, Dreon2017oca} and which also offers interesting possibilities for optical dipole trapping. A particular motivation for the experiments pursued in our laboratory is the exciting prospect to realize novel superfluid states in mass-imbalanced fermion mixtures \cite{Gubbels2013ifg, Wang2017eeo, Pini2021bmf}, which is the reason why we work with the fermionic isotopes $^{161}$Dy and $^{40}$K.

Our work is structured as follows. In Sec.~\ref{sec:methods}, we describe the experimental procedures, including the preparation protocol and probing methods of the ultracold gas in the optical lattice. In Sec.~\ref{sec:results}, we discuss our main results on the dynamical polarizability of dysprosium for varying optical detunings and polarizations near the 626-nm line. We then extract the contribution of scalar and tensorial components and obtain the linewidth of the transition. In addition, we demonstrate dipole trapping and measure the heating rate and lifetime in Sec.~\ref{sec:dipoltrap}. In Sec.~\ref{sec:conclusion}, we finally summarize our results and give a brief outlook.

\section{Methods}\label{sec:methods}
In this section, we present the methods used to determine the dynamical polarizability of Dy near the 626-nm line. We start by summarizing the experimental sequence to obtain an ultracold sample of either Dy or K atoms in the lattice (Sec.~\ref{sec:sampleprep}), after which we describe the methods to measure the lattice depth for the two species (Sec.~\ref{sec:Dymethod}) and how we use K as a well-known reference to calibrate our measurement on Dy (Sec.~\ref{sec:Kmethod}).

\begin{figure}[b]
\includegraphics[width=\columnwidth]{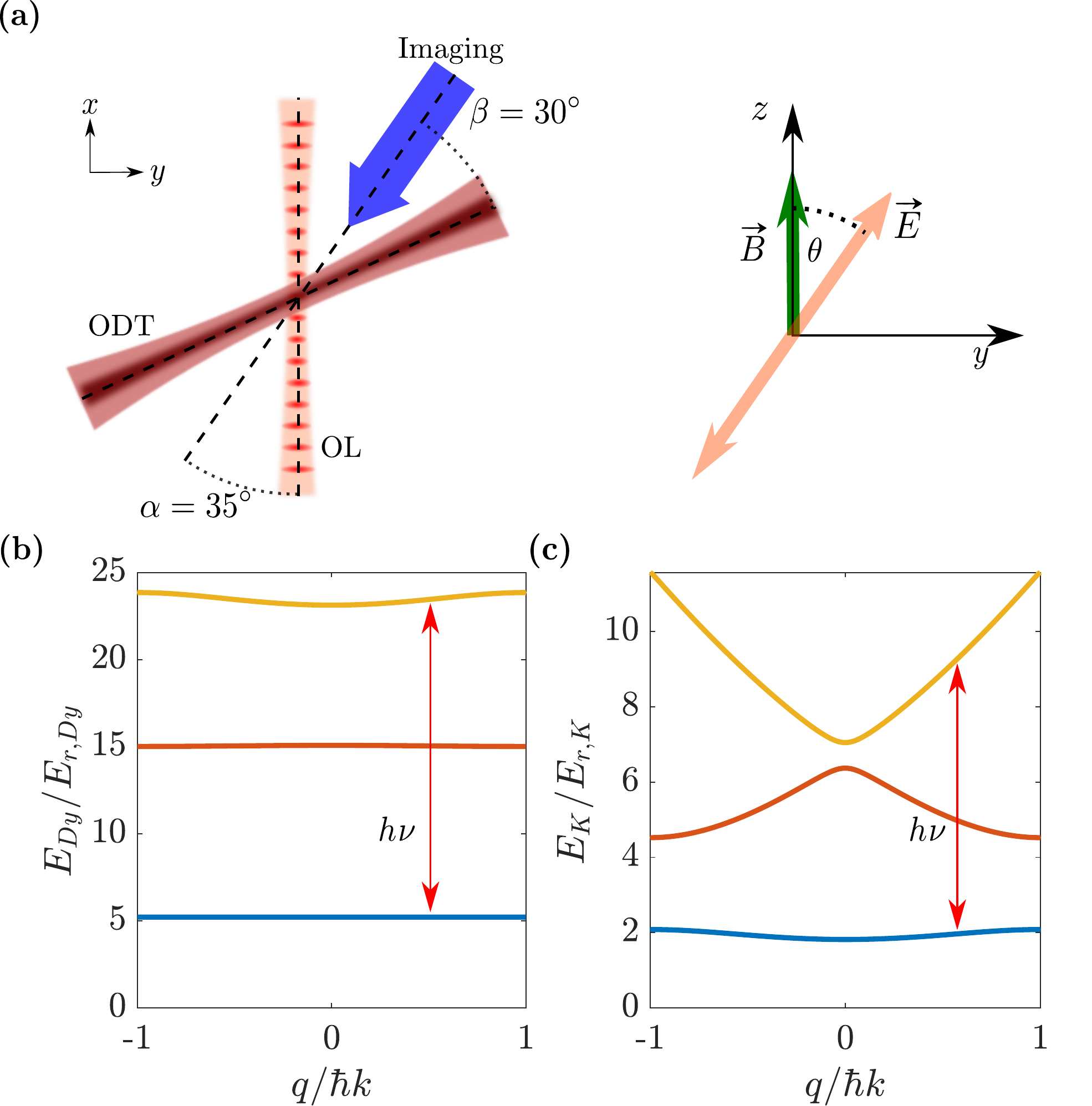}
\caption{\label{fig:Setup} (a) Schematic of our experiment. The 1D optical lattice beam (OL) is overlapped by the optical dipole trap (ODT) and imaging beam at the sample position. The second beam of the crossed dipole trap propagates along the $z$ direction (direction of gravity) and is not shown here. The external magnetic field $\vec{B}$ is aligned with the gravity axis, defining the polarization angle $\theta$ for the $\vec{E}$ field of the lattice beam. $\theta$ can be changed by rotating a half-wave plate. (b) and (c) Band structures of an optical lattice for Dy and K with a typical depth of $V_{Dy}=30 E_{r,Dy}$ and $V_K=5 E_{r,K}$, respectively. Atoms are transferred if the photon energy $h\nu$ from the modulation matches the energy difference between bands at a given quasimomentum $q$.}
\end{figure}

\subsection{Sample preparation} \label{sec:sampleprep}
Our experiments begin with preparing spin-polarized degenerate Fermi gases of $^{161}$Dy or $^{40}$K in an optical dipole trap, following procedures described in our previous work~\cite{Ravensbergen2018poa}. For Dy, we rely on the evaporation of atoms in a single spin state in a crossed optical dipole trap, taking advantage of universal dipolar collisions~\cite{Bohn2009qud}. At the end of the evaporation, we are left with a typical atom number of $N_\mathrm{Dy}=2\times10^4$ in the absolute ground state $|F = 21/2,m_F=-21/2\rangle$. The mean (geometrically averaged) trapping frequency is $\bar{\omega}_\mathrm{Dy}/(2\pi)=120$~Hz, and the sample is at a temperature of $T/T_\mathrm{F,Dy}=0.1$, where $T_\mathrm{F,Dy}$ is the Fermi temperature of the trapped sample. 

To produce degenerate samples of K, we load Dy and K together in the crossed dipole trap. Since the trap is about 3.6 times deeper for K than for Dy~\cite{Ravensbergen2018ado} and the sample is nearly thermalized, essentially Dy atoms get lost during evaporation, and K is sympathetically cooled by Dy. After fully evaporating all remaining Dy atoms, we end up with a pure sample of $^{40}$K in the ground state $|F=9,m_F=-9/2\rangle$, with $N_\mathrm{K}=1\times10^4$, $T/T_\mathrm{F,K}=0.2$, and $\bar{\omega}_\mathrm{K}/(2\pi)=450$~Hz.

The atoms are then adiabatically loaded into a one-dimensional (1D) optical lattice generated by two counterpropagating, linearly polarized laser beams at wavelength $\lambda \approx626$~nm with a beam waist $w_0 = 55~\mu$m and a power $P$ in the range between 17 and 200~mW per beam. For normalization purposes, we define a reference power of $P_0 = 67$~mW. The lattice is superimposed with the crossed dipole trap used for evaporation [see Fig.~\ref{fig:Setup}(a)]. The lattice beams are oriented horizontally, and the quantization axis is defined by applying a small magnetic field less than 1~G along the direction of gravity. We ramp up the lattice potential
\begin{equation}
    V_i(\textbf{r},\omega)=-\frac{2\pi a_0^3}{c}\tilde{\alpha}_i(\omega) I(\textbf{r}),
\end{equation}
where $\omega$ is the laser frequency, $I(\textbf{r})$ is the laser intensity at position $\textbf{r}$, $a_0$ is the Bohr radius, and $c$ is the speed of light, in 200~ms to a certain lattice depth $\hat{V}$.
As in our previous work~\cite{Ravensbergen2018ado}, we define $\tilde{\alpha}_i(\omega)$ as the dimensionless real part of the dynamical polarizability of atomic species $i\in\{\mathrm{Dy,K}\}$ normalized to the atomic unit of polarizability. The optical lattice depth is typically expressed in units of recoil energies $E_{r,i}=h^2 / (2m_i \lambda^2)$, where $h$ is the Planck constant and $m_i$ is the atomic mass.
After loading, because of the deeply degenerate nature of the samples, the atoms completely fill the ground band, and the fractional population of the atoms in the excited bands is measured to less than 6\%, which we verified by a band mapping technique~\cite{Kastberg1995aco,Greiner2001epc}. We verified that ramping up the lattice intensity and then ramping it down again are possible without significant heating of the samples.

To mitigate the antitrapping effect when working with blue-detuned lattices, we ramp up the dipole trap power simultaneously with the lattice to a trapping frequency of $\bar{\omega}_\mathrm{Dy}/(2\pi)=190$~Hz and $\bar{\omega}_\mathrm{K}/(2\pi)=700$~Hz. This also helps us reduce the differential gravitational sag that the two species experience, which would result in a difference of about $5~\mu$m in the vertical direction and therefore also a difference in lattice intensity experienced by the atoms. The deeper trap reduces the differential sag to about $1~\mu$m.

\subsection{Measuring the lattice depth of $^{161}$Dy}\label{sec:Dymethod}
In order to determine the lattice depth of $^{161}$Dy atoms, we perform amplitude-modulation spectroscopy by sinusoidally varying the depth of the optical lattice potential for $100$ to $200$~ms with a relative amplitude of about $5$\%. In this method, the population initially filling the ground band ($n=0$) is excited to the higher bands by absorbing the photons resonant to the energy difference between bands [see Fig.~\ref{fig:Setup}(b)]. 
Because of the curvature of the bands, only a specific class of quasimomenta $q$ is resonant with the excitation frequency and can be transferred as the modulation frequency is swept~\cite{Heinze2011mso}. The amplitude-modulation scheme predominantly drives $\Delta n=2$ excitations because of parity conservation, coupling the ground and second excited bands with frequency $\nu$~\cite{Denschlag2002abe}. 

The superimposed dipole trap mixes all spatial dimensions so that transitions to higher bands result in heating of the sample caused by the momentum being added. The transition probability is dependent on $q$ and has a maximum at the lower band edge, where $q=0$, and drops for larger $q$. We therefore expect a sharp increase in cloud size when the modulation frequency matches the resonance condition, $E_{n,q}-E_{m,q} = h\nu$ at $q=0$. Here, $E_{n,q}$ is the energy of a particle in the $n$th band with quasimomentum $q$.
To observe this effect, we ramp down the lattice (in about 2~ms), then switch off the dipole trap and measure the size of the atomic cloud using standard absorption imaging after typically $5$~ms of free expansion. We then determine the size of the atomic cloud $\sigma$ using a Gaussian fit.

\begin{figure}[t]
\includegraphics[width=\columnwidth]{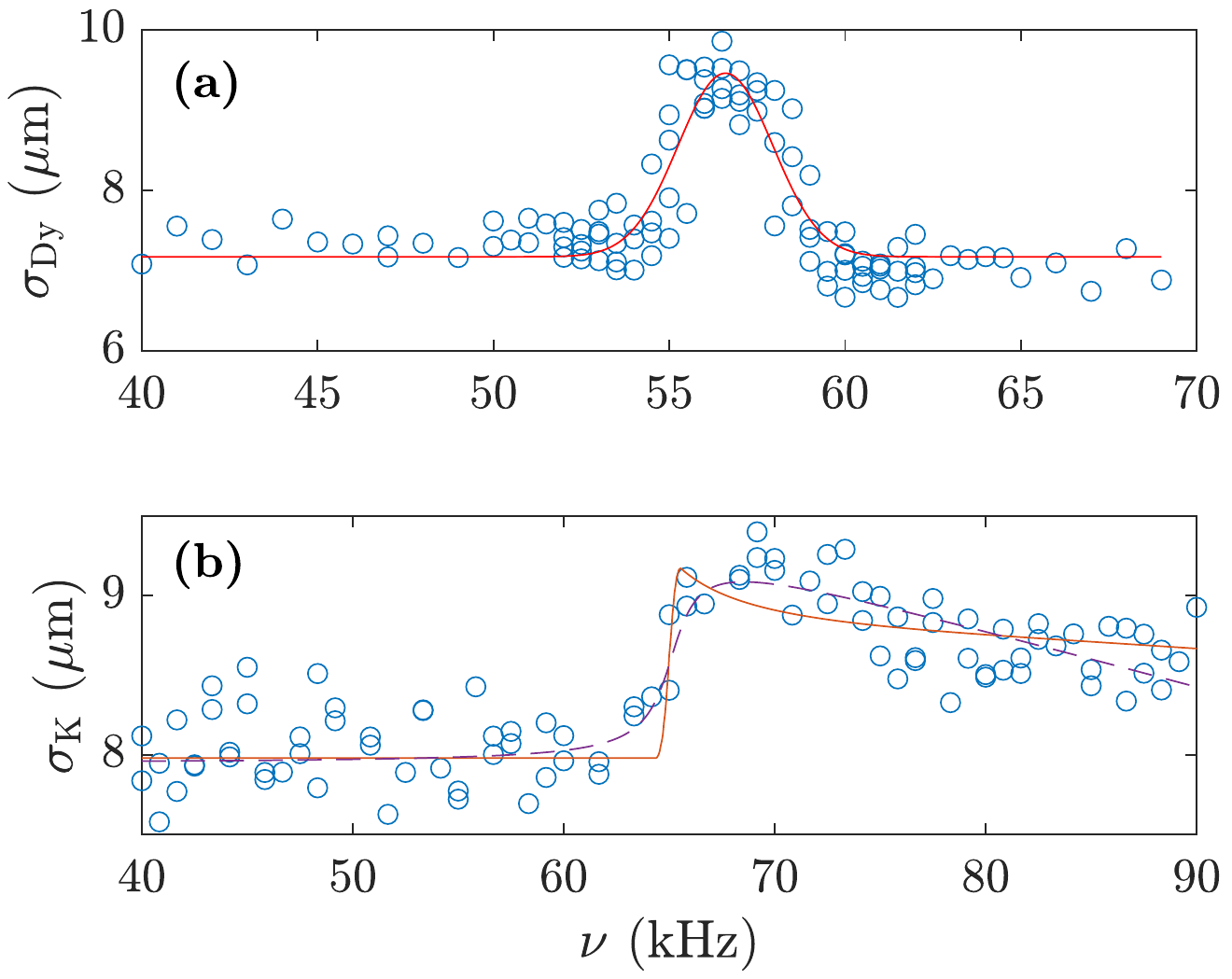}
\caption{\label{fig:DyKexcombined}Representative amplitude-modulation spectra for an optical lattice at $\lambda=626.174$~nm. The atomic cloud size is plotted as a function of the modulation frequency. (a) For $^{161}$Dy, at a power $P = 33.5$~mW, we obtain a lattice depth of $\hat{V}_\mathrm{Dy} = 28.96(9) E_\mathrm{r,Dy}$ by matching $\nu_0-\Delta\nu$, extracted from a Gaussian fit (solid line), to the band gap. (b) For $^{40}$K the spectrum, taken at $P = 67$~mw, shows an asymmetric profile due to the broad band structure. We take into account systematic errors in the numerical simulation used to determine the lattice depth. The fitted simulation (solid line) yields a depth of $\hat{V}_\mathrm{K} = 4.70(14) E_\mathrm{r,K}$, which is consistent with the depth of $\hat{V}_\mathrm{K} = 4.73(8) E_\mathrm{r,K}$ obtained from an arctangent fit (dashed line). The errors given here represent the statistical fitting errors. The systematic errors are much larger (see text).}
\end{figure}

Figure~\ref{fig:DyKexcombined}(a) shows a typical amplitude-modulation spectrum for the $^{161}$Dy atoms, plotted as a function of the modulation frequency. The spectrum is fitted with a Gaussian function $\sigma(\nu) \propto e^{-((\nu-\nu_0)^2/2\Delta \nu^2)}$, where $\nu_0$ and $\Delta \nu$ are fitting parameters, indicating the frequency position of the lower band edge for the given lattice depth. For each choice of the wavelength $\lambda$, the power of the lattice beams is set such that the lattice for $^{161}$Dy is deep enough (more than $25 E_{r,\mathrm{Dy}}$) to generate flat bands. Close to resonance, the power is kept low enough to avoid heating by photon scattering. This narrows the spectroscopy signal and allows the use of a Gaussian fitting function.
The typical width of the $n=2$ band in this regime is less than 7\% of the gap between the two bands. We obtain the depth of our optical lattice $\hat{V}_\mathrm{Dy}$ by matching $\nu_0-\Delta \nu$ with the lower band edge calculated by a band structure model for an infinite, homogeneous one-dimensional lattice. We define \begin{equation}
    s_\mathrm{Dy}(\lambda)  = \frac{\hat{V}_\mathrm{Dy}(\lambda,P)}{E_\mathrm{r,Dy}} \frac{P_0}{P}
\end{equation}
as the power-normalized lattice depth in units of $E_{r,i}$. Here, the power normalization scales the lattice depth to the reference value $P_0$. For the example in Fig.~\ref{fig:DyKexcombined}(a), we obtain $s_\mathrm{Dy} = 57.9(2)$ at $\lambda=626.174$~nm. The uncertainty given here represents the statistical fitting error.

\subsection{Calibration measurements with potassium}\label{sec:Kmethod}
For calibration purposes we perform a similar lattice depth measurement with $^{40}$K. After preparing the K sample in the lattice, we modulate the amplitude of the lattice beam for $500$~ms and image the atoms after $2$~ms of time of flight. The laser beam at $\lambda=626$~nm is far detuned from the potassium transition lines, resulting in an accurately known polarizability value of $\tilde{\alpha}_\mathrm{K}=-556(1)$ as a reliable reference with negligible anisotropic contributions~\cite{Safronova2013mwf}. We checked that the K lattice depth depends neither on the particular wavelength chosen close to the Dy resonance line nor on the polarization angle. We also verified the expected linear scaling of the lattice depth with the lattice power in a range between $P_0$ and $3 P_0$.

Figure~\ref{fig:DyKexcombined}(b) exhibits an example of an amplitude-modulation spectrum for $^{40}$K at the reference power $P_0 = 67$~mW. The cloud size as a function of $\nu$ shows a pronounced asymmetry, pointing to the band edge near $66$~kHz. Because of the relatively small value of $\tilde{\alpha}_\mathrm{K}$ at $\lambda=626$~nm and the large recoil energy $E_\mathrm{r,K}\approx 4E_\mathrm{r,Dy}$, the lattice depth for the potassium atoms becomes small, which leads to a broad band structure and therefore to a broader spectral response. The spectrum is further broadened for various technical reasons~\cite{TechnicalReasonsNote}, which makes the identification of the exact location of the band gap at $q = 0$ difficult (see Appendix~\ref{sec:appendix}). To analyze the spectra, we use a combination of analytical fitting functions and a numerical simulation based on $q$-dependent transition probabilities calculated between the ground and second excited bands. The uncertainty in the identification of the band edge leads to a systematic error, which we estimate to be $4\%$. With this model, we deduce a $^{40}$K lattice depth of $s_\mathrm{K}=4.75(19)$ for the same conditions as used in the $^{161}$Dy measurements. 

Finally, the dynamical polarizability of the dysprosium atoms can be derived as
\begin{equation}
    \tilde{\alpha}_\mathrm{Dy}(\lambda) =
    \frac{s_\mathrm{Dy}(\lambda)}{s_\mathrm{K}}\frac{m_\mathrm{K}}{m_\mathrm{Dy}}\tilde{\alpha}_\mathrm{K},
\end{equation}
which is the basis of our further analysis. While the main uncertainty arises from the determination of the band edge, we have identified a second possible source of systematic uncertainty. The spatial distributions of both species in the optical lattice may differ, which leads to slightly different sample-averaged lattice depths. We have modeled this effect by employing the same numerical simulation as used for K also for Dy for a range of different experimental parameters. For the effect of the spatial distribution on the determination of $\tilde{\alpha}_\mathrm{Dy}$, we estimate a systematic error of $3\%$, which together with the band-edge uncertainty of $4\%$ adds up to a total combined systematic error of $5\%$.

\section{Results}\label{sec:results}
In this section we present the main results of our measurements of the anisotropic polarizability of Dy and its variation with detuning across the 626-nm resonance. Furthermore, we extract the natural linewidth of the transition.
\subsection{Anisotropic polarizability}\label{sec:aniso}

\begin{figure}[]
\includegraphics[width=\columnwidth]{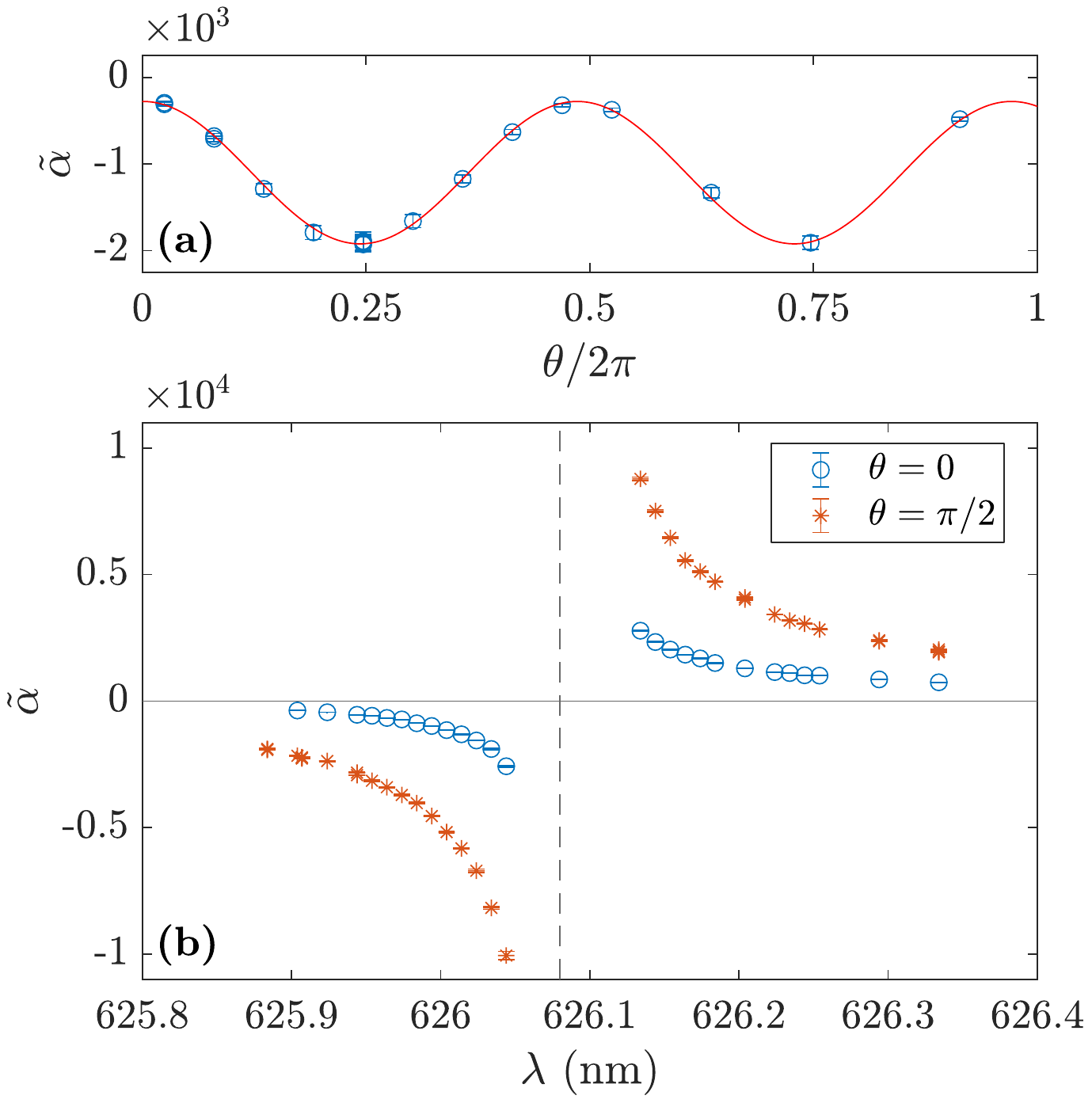}
\caption{\label{fig:Dyanglewlcombined} Measurements of anisotropic polarizability for $^{161}$Dy. (a) Angle dependence of the polarizability at $\lambda=625.884$~nm. The variation reveals the scalar and tensor contributions. The solid line shows a fit according to Eq.~(\ref{eq:STpol}). (b) Wavelength dependence of the dynamical polarizability of $^{161}$Dy near the intercombination transition line for two different polarization angles $\theta = 0,\pi/2$ (parallel and perpendicular to the quantization axis). The dashed line indicates the resonance center. In (a) the error bars represent the $1\sigma$ statistical fitting errors from the individual spectra used to determine the lattice depth. In (b) the error bars are smaller than the symbol size.}
\end{figure}

The dynamical polarizability can be generally decomposed into scalar, vector, and tensor parts~\cite{Deutsch2010qca,LeKien2013dpo}, which we denote $\tilde{\alpha}_s$, $\tilde{\alpha}_v$, and $\tilde{\alpha}_t$, respectively. The present work employs linearly polarized lattice beams and, consequently, measures the scalar and tensor contributions. The dynamical polarizability of an atom in the stretched state can be expressed as a weighted sum of scalar and tensor components,

\begin{equation}
\label{eq:STpol}
    \tilde{\alpha} (\omega) = \tilde{\alpha}_s (\omega) + \frac{3\cos^2\theta-1}{2} \tilde{\alpha}_t (\omega),
\end{equation}
where $\theta$ is the polarization angle defined with respect to the quantization axis (see Fig.~\ref{fig:Setup}) and $\omega=2\pi c/\lambda$ is the angular frequency of the laser field. In the experiment, we scan $\theta$ by rotating a half-wave plate. The quantization axis is determined by applying a small magnetic field less than 1~G along the direction of gravity (see Fig.~\ref{fig:Setup}). We experimentally confirmed that our measurements remain unaffected by an external magnetic field up to 10~G. 
In Fig.~\ref{fig:Dyanglewlcombined}(a), we plot the dynamical polarizability as a function of the angle $\theta$. The measurement was carried out at a fixed wavelength of $\lambda=625.884$~nm, and the value of $\tilde{\alpha}_\mathrm{Dy}$ is derived from the lattice depth, as discussed before. The variation of $\tilde{\alpha}_\mathrm{Dy}$ shows the expected mixing of the scalar and tensor polarizability, depending on $\theta$. A fit according to Eq.~(\ref{eq:STpol}) gives $\tilde{\alpha}_s=-1.37(1)\times10^3$ and $\tilde{\alpha}_t=1.10(1)\times10^3$ ($1\sigma$ statistical fitting errors). Here, at the specific wavelength chosen, the tensor component provides a contribution to the total dynamical polarizability that is comparable to the scalar component, generating the ratio $\tilde{\alpha}_t/\tilde{\alpha}_s= -0.80(1)$. Figure~\ref{fig:Dyanglewlcombined}(b) shows the total polarizability for the two angles, $\theta = 0$ and $\pi/2$, from which we obtain $\tilde{\alpha}_s$ and $\tilde{\alpha}_t$ according to Eq.~(\ref{eq:STpol}). We repeat the measurements for various detunings and observe the variation of the absolute value of $\tilde{\alpha}_\mathrm{Dy}$ in a region of roughly $0.5$~nm around the resonance center. The sign follows from the fact that the light field is attractive ($\tilde{\alpha}_s > 0$) for red detuning and changes sign on the blue side of the resonance.

\begin{figure}[t]
\includegraphics[width=\columnwidth]{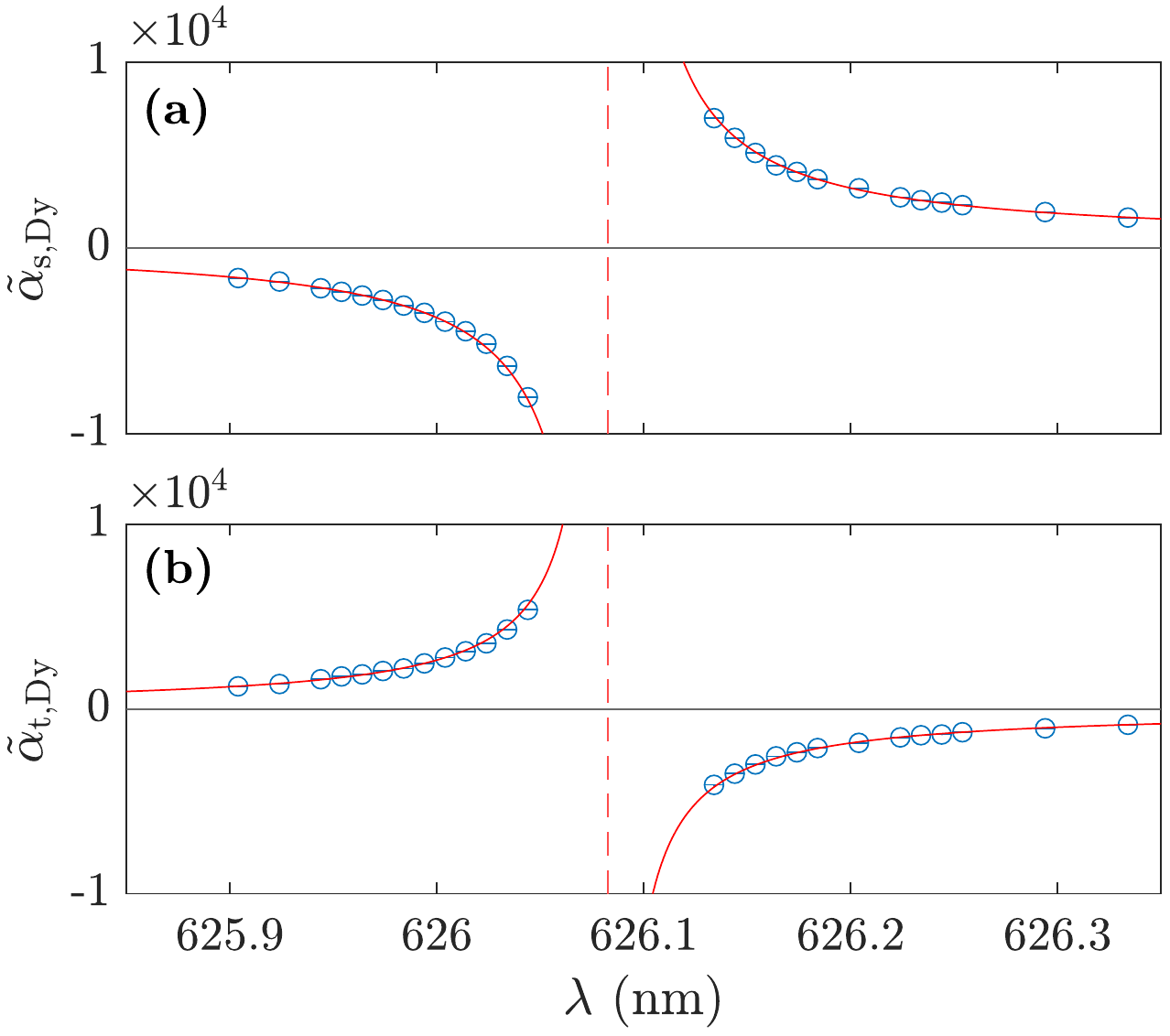}
\caption{\label{fig:DySkalTensCombined} Measured (a) scalar and (b) tensor components of the dynamical polarizability of $^{161}$Dy near the 626-nm line. Solid lines show a fit according to Eq.~(\ref{eq:resmodel}). Error bars are smaller than the symbol size.}
\end{figure}

Figure~\ref{fig:DySkalTensCombined} shows the final result for the resonance behavior of $\tilde{\alpha}_s$ and $\tilde{\alpha}_t$. The polarizability can be modeled with a resonance model
\begin{equation}
\label{eq:resmodel}
    \tilde{\alpha}(\omega) =\tilde{\alpha}_{bg} + \beta\frac{\omega_{0}^2}{\omega_{0}^2-\omega^2},
\end{equation}
where $\tilde{\alpha}_{bg}$ is a background contribution from other resonances, the parameter $\beta$ is defined as a dimensionless resonance strength, and $\omega_0 = 2\pi c/\lambda_0$ is the resonance center angular frequency. This model is applied to the data for both $\tilde{\alpha}_s$ and $\tilde{\alpha}_t$. In this case, we fit the data with a single-resonance model, although three hyperfine resonances are actually present in the fermionic isotope in the stretched state. Since the hyperfine splitting is small compared to the detuning~\cite{Eliel1980aso2,Lu2011soa,Dreon2017dis}, the deviation from the single-resonance model is negligible compared to our experimental uncertainties. We confirmed this by fitting the data with a corresponding extended model that takes hyperfine resonances into account.

Table~\ref{tab:fitresults} summarizes the fitting results. Notably, $\tilde{\alpha}_s$ includes a background of 275(13), originating from other transitions, mostly the strong blue line near $421~$nm. In contrast, the background in $\tilde{\alpha}_t$ is only 8(14), which is consistent with 0. The off-resonant contributions from other lines essentially cancel each other in the tensorial part. For the 626-nm transition, from theory describing the angular part of a $J=8\rightarrow J' =9$ transition~\cite{LeKien2013dpo,Safronova2013mwf,Li2017aot}, we expect a ratio between the tensor and scalar parts on resonance of
\begin{equation}
    r \equiv \lim_{\omega\rightarrow\omega_0} \frac{\tilde{\alpha}_t(\omega)}{\tilde{\alpha}_s(\omega)} = \frac{\beta_t}{\beta_s}= -40/57\approx-0.7018.
\end{equation}
However, fitting the data with Eq.~(\ref{eq:resmodel}) yields a ratio of $-0.643(4)$. We attribute this deviation to a systematic error resulting from setting $\theta$ in our measurements (see Appendix~\ref{sec:appendix_B}). We note that the fit results for the resonance position are inconsistent within the very small fit uncertainties. We attribute this minor discrepancy to the fact that we model the contribution of other lines with a simple constant offset $\tilde{\alpha}_{bg}$, thus ignoring the effect of a possible slope in the background. The exact value resulting for the fit parameter $\omega_0 = c/\lambda_0$ may be sensitive to such a slope. However, this minor inconsistency has no significant effect on the values obtained for the resonance strength parameter $\beta$.

\begin{table}[]
\caption{Results for the resonance parameters $\tilde{\alpha}_{bg}$, $\beta$, and $\lambda_0 = c/\omega_0$, obtained by fitting Eq.~(\ref{eq:resmodel}) to our data sets for $\tilde{\alpha}_{s}(\omega)$, $\tilde{\alpha}_{t}(\omega)$, and the mean polarizability $\tilde{\alpha}_{0}(\omega)$ according to Eq.~(\ref{eq:meanpol}). Numbers in parentheses give the $1\sigma$ fit uncertainties.}\label{tab:fitresults}
\begin{tabular}{c|rrr}
Data                 & $\tilde{\alpha}_{bg}$ & $\beta$ & $\lambda_0$~(nm)  \\ \hline
$\tilde{\alpha}_{s}$ & 275(13)               & $1.055(3)$         & 626.0808(5) \\
$\tilde{\alpha}_{t}$ & 8(14)                 & -0.679(4)        & 626.0794(8) \\
$\tilde{\alpha}_{0}$ & 278(10)               & 0.885(3)        & 626.0850(4)
\end{tabular}
\end{table}

\subsection{Determination of the natural linewidth}
To avoid the effect of uncertainties in $\theta$ we introduce the mean polarizability
\begin{equation} \label{eq:meanpol}
    \tilde{\alpha}_0 = \frac{\tilde{\alpha}(\theta=0)+\tilde{\alpha}(\theta=\pi/2)}{2} = \tilde{\alpha}_s+\frac{1}{4}\tilde{\alpha}_t,
\end{equation}
which turns out to be insensitive to small deviations of $\theta$ from the ideal values $0$ and $\pi/2$ (see Appendix~\ref{sec:appendix_B}). We can fit $\tilde{\alpha}_0$ with the model introduced in Eq.~(\ref{eq:resmodel}); the results can again be found in Table~\ref{tab:fitresults}. Notably, we extract $\tilde{\alpha}_{bg} = 277(13)$, which is consistent with the offset on the scalar component given before. With the definition of $\tilde{\alpha}_0$ we find the relation
\begin{equation} \label{eq:resmodel0}
     \beta_0 = \left(1+\frac{r}{4}\right)\beta_s,
\end{equation}
which now includes the ratio $r$, fixed to a theoretical value of $-0.7018$. Our result for the resonance strength $\beta=0.885(3)$ is now insensitive to systematic errors in the angle determination and combines both sets of data for $\theta = 0$ and $\pi/2$. With this method, we are left with the dominant error being the $5\%$ uncertainty in the calibration of $\tilde{\alpha}_\mathrm{Dy}$ as discussed before.

We can now extract the natural linewidth
\begin{equation}
    \Gamma = \frac{2 a_0^3 \omega_0^4}{c^3}\frac{2J+1}{2J'+1}\frac{\beta_0}{1+r/4}
\end{equation}
of the closed $J=8\rightarrow J'=9$ transition. We calculate a linewidth of $\Gamma/2\pi =(137.9 \pm0.4_{\text{stat}} \pm 6.9_{\text{sys}})$~kHz, which agrees well with transition probabilities obtained by radiative lifetime measurements on atomic beams~\cite{Gustavsson1979lmf,Curry1997lif,Wickliffe2000atp}. The relative uncertainty is on par with the most precise measurement of the lifetime of $1.17(3)~\mu$s~\cite{Gustavsson1979lmf}, which corresponds to a natural linewidth of $(136\pm3)$~kHz. The agreement of our result with this benchmark of a direct lifetime measurement on the level of a few percent also confirms that our indirect way to determine line strengths via measurements of dynamic polarizabilities produces accurate results. With an optimization of experimental parameters, the uncertainty in the determination of the lattice depth of K, which is the source of the dominating systematic error, could be reduced further.

\section{Demonstration of optical dipole trapping} \label{sec:dipoltrap}
In an additional experiment, we realize an optical dipole trap operating on the 626-nm transition and measure the lifetime and heating rate of the dysprosium atoms. We set the laser wavelength to the red-detuned side of the resonance and use one of the lattice beams, with the counterpropagating beam blocked. The polarization angle is set to $\theta=\pi/2$ to maximize the polarizability. By slowly (within 100~ms) ramping down the power of the horizontal 1064-nm dipole trap beam, we load the atoms into a bichromatic trap consisting of the horizontal 626-nm beam and the vertical 1064-nm dipole trap beam, with an average trapping frequency of $\bar{\omega}/(2\pi)=110$~Hz. After a variable hold time, we record atom number and temperature with standard time-of-flight imaging. Since the lifetime in the $1064$-nm dipole trap is two orders of magnitude larger than any other timescale of the system, we consider only the heating effect originating from the $626$-nm trap. 

\begin{figure}[b]
\includegraphics[width=\columnwidth]{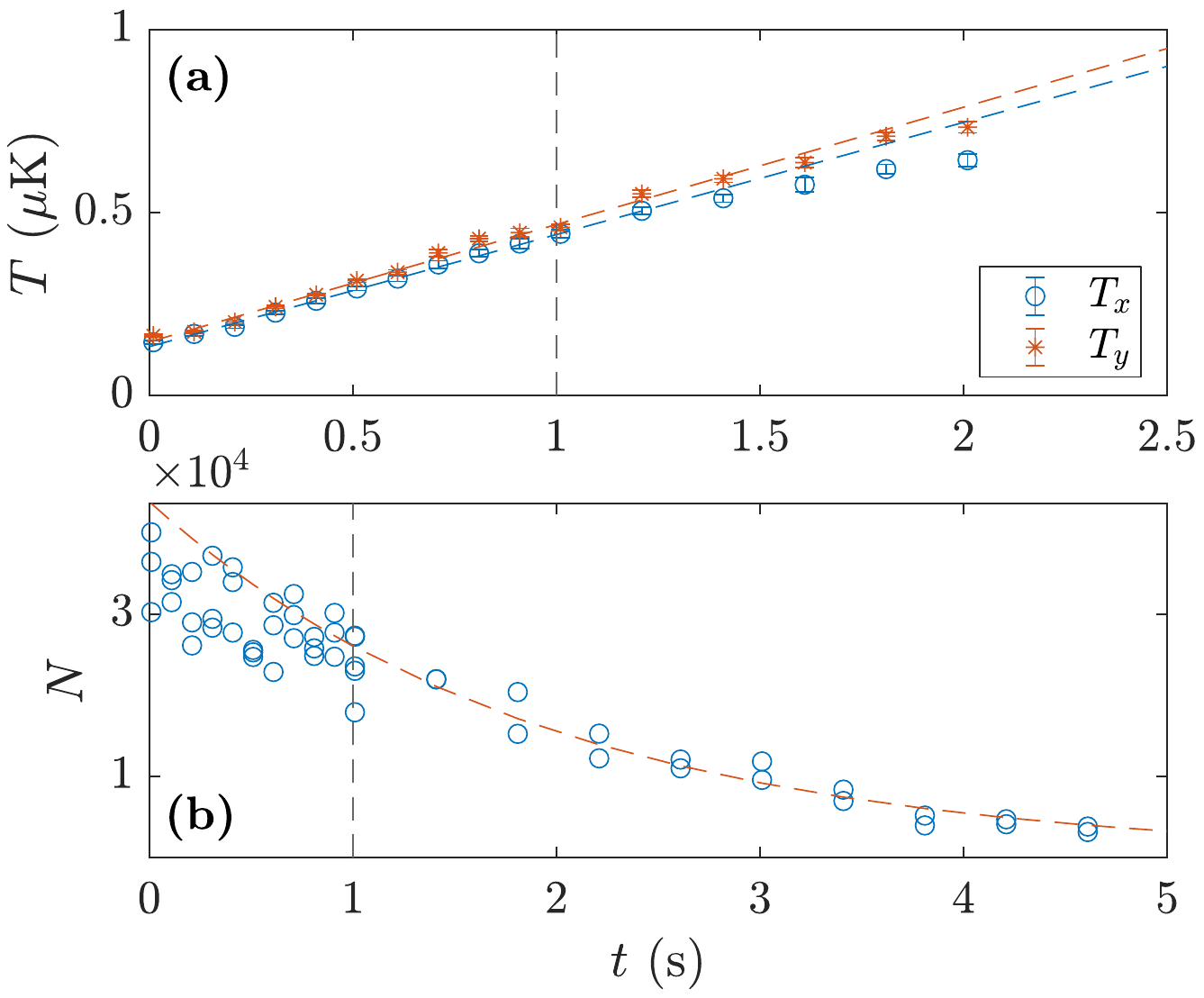}
\caption{\label{fig:DyTrap} Time evolution of temperature and atom number in a bichromatic crossed dipole trap. (a) Temperature in the $x$ and $y$ directions with linear fits for the first 1~s (dashed vertical line). (b) Atom number and corresponding exponential fit from 1~s onward.}
\end{figure}

In Fig.~\ref{fig:DyTrap}, the time evolution of the temperature and atom number at $\lambda=626.334$~nm are displayed. At this detuning, we measured the polarizability to be $\tilde{\alpha} = 1.97(8)\times10^3$, which includes also a possible deviation from the ideal angle $\theta = \pi/2$. In the measurement, we ramp the laser beam power to 170~mW, which results in a central intensity of $I_0 = 3.5\times10^6~\text{mW~cm}^{-2}$. We calculate a central optical potential depth of $U_0 = -2\pi a_0^3 \tilde{\alpha} I_0/c =  k_B\times16~\mu$K, and by taking the gravitational effect into account, the potential depth is reduced to $k_B\times4\mu$K. Initially, we observe a linear increase of the temperature. A linear fit from $0$ to $1$~s yields a heating rate of $311(7)$~nK/s, which indicates a photon scattering rate of about $0.8~\text{s}^{-1}$. The calculated photon scattering rate in the middle of the trap is~\cite{Grimm2000odt}
\begin{equation}
    R_\mathrm{scatt}= \frac{\Gamma}{\hbar\Delta }U_0,
\end{equation}
where we take our result for the linewidth $\Gamma=2\pi\times 138(7)$~kHz and where $\Delta = \omega-\omega_0$ is the frequency detuning. For our experimental parameters, we calculate a scattering rate of $R_\mathrm{scatt}\approx1.5$ $\mathrm{s}^{-1}$ in the center of the optical potential. However, this model neglects that the atoms are spatially distributed in the trap and sample areas with lower intensity than in the trap center. This effect is even enhanced by the influence of gravity, which shifts the trap center out of the center of the intensity distribution. Furthermore, there is a considerable uncertainty in the measurement of the beam waist and therefore the value of $I_0$. Considering these effects, the observed heating is consistent with the expected photon scattering.

For longer hold times, the heating rate is observed to decrease. This might be because the increased cloud size leads to a lower average intensity across the sample and therefore a reduced scattering rate. Another explanation is that when the temperature reaches about 500~nK, which is about a factor of 8 below the trap depth, some evaporation may set in and counteract the heating. Indeed, we observe an increased atom loss rate after 1~s of hold time [see Fig.~\ref{fig:DyTrap}(b)]. We use an exponential fit from 1~s onward and obtain a lifetime of $\tau=1.9(1)$~s. 

The measurement shows that dipole trapping close to the 626-nm line with a rather small wavelength detuning works as expected and provides an additional versatile tool to tailor optical potentials for Dy atoms. In particular, this may be interesting for species-selective dipole traps to manipulate mixtures of Dy with other species and can be used to optimize conditions to obtain superfluid regimes in Fermi-Fermi mixtures~\cite{Ravensbergen2020rif,Pini2021bmf}.

\section{Conclusion and outlook}\label{sec:conclusion}
We have shown that the method introduced in Ref.~\cite{Ravensbergen2018ado} to accurately measure the dynamic polarizability of an atom by comparison with a reference species can be generalized to light fields that act repulsively. Using modulation spectroscopy in an optical lattice, we investigated the 626-nm intercombination line of Dy and measured the scalar and the tensorial part of the anisotropic polarizability in the resonance region. As an important benchmark for our method, the line strength derived from our polarizability measurements is consistent with previous measurements of the natural transition linewidth. Our relative uncertainty of $\sim$5\% is already on par with the previous measurements and may be further improved by further suppressing systematic effects. The method is of particular interest for characterizing the multitude of optical transitions in submerged-shell lanthanide atoms, which have become very popular in laser cooling and quantum gas experiments.

We have also demonstrated optical dipole trapping of Dy with laser light tuned about 0.25\,nm below the center of the 626-nm line. We found efficient trapping with low heating, in quantitative agreement with expectations based on the line strength derived from the polarizability measurements. This introduces optical dipole potentials generated by laser light tuned close to this intercombination line as an interesting tool for future experiments.

For our particular goal to create a mass-imbalanced fermionic superfluid in the $^{161}$Dy-$^{40}$K mixture \cite{Ravensbergen2018poa, Ravensbergen2020rif}, species-specific optical potentials \cite{Leblanc2007sso} offer alternative handles for control. On the blue side of the 626-nm resonance, the light field will be repulsive for both species. This allows us to create boxlike trapping schemes \cite{Ovchinnikov1997stf, Gaunt2013bec, Mukherjee2017haf} for the preparation of homogeneous Dy-K mixtures. At a specific detuning, the polarizability ratio will match the mass ratio, and an optical levitation scheme \cite{Shibata2020cog} can be realized that compensates the effect of gravity for both species simultaneously. In a harmonic trap, the phase diagram critically depends on the trap frequency ratio of both species, as investigated theoretically in Ref.~\cite{Pini2021bmf}. Species-specific optical potentials allow us to optimize the conditions for attaining and observing the superfluid phase transition.

\begin{acknowledgments}

We thank M.\ Lepers for discussion. We acknowledge financial support from the Austrian Science Fund (FWF) within Project No.\ P32153-N36 and within the Doktoratskolleg ALM (Grant No.\ W1259-N27). We further acknowledge a Marie Sklodowska Curie fellowship awarded to J.H.H.\ by the European Union (project SIMIS, Grant Agreement No. 894429).

\end{acknowledgments}


%


\appendix
\section{Lattice depth extraction for potassium}\label{sec:appendix}
In general, the Hamiltonian of a lattice modulated with modulation frequency $\nu$ is given by
\begin{equation}
\label{eq:latham}
    H = \frac{p^2}{2m} + V \cos (k_x x)^2 \left[1 + \epsilon \cos(2\pi\nu t)\right],
\end{equation}
where $\epsilon$ denotes a small perturbation of the lattice depth and $k_x = 2\pi/\lambda$ is the wavenumber of the laser in the $x$ direction. The calculation of transition probabilities between bands of this lattice follows Ref.~\cite{Hundt2011dipl}. According to Bloch's theorem, the eigenstates of the unperturbed system can be described in a plane-wave basis by
\begin{equation}
    \Psi_q^{(n)}(x) = e^{iqx} \sum_K c_{K,q}^{(n)} e^{iKx},
\end{equation}
where $c_{K,q}^{(n)}$ are the Fourier coefficients to the reciprocal lattice vectors $K$ in the $n$-th band. Using Fermi's golden rule, the transition probability between bands $n$ and $m$ of an atom with quasimomentum $q$ is given by
\begin{equation}\label{eq:transprob}
    W_q^{nm} \propto |\sum_K c_{K,q}^{(n)} c_{K,q}^{(m)} (q + K)^2 |^2.
\end{equation}
By numerically diagonalizing the Hamiltonian in Eq.~(\ref{eq:latham}), the coefficients $c_{K,q}^{(n)}$ can be found for all available reciprocal lattice vectors $K$, the energy gap between bands $n$ and $m$ can be calculated for a given $q$, and $W_q^{nm}$ can be converted to $W^{nm}(\nu)$. For the transition between bands $n=0 \rightarrow m=2$, the resulting spectrum $W^{02}(\nu)$ exhibits a sharp edge on the lower-energy side, which corresponds to atoms with $q=0$. However, if the cloud width $\sigma_c$ and lattice beam waist $w_0$ are comparable and if the cloud center position is offset from the center of the lattice by $r_c$, the effective lattice depth will vary over the extent of the cloud, effectively smoothing out the sharp edge. In a numerical simulation we account for this by slicing the atom distribution and calculating the transition probability with the corresponding $V(r)$ for each slice, where $V(r)$ follows the Gaussian form of the lattice beam. Each spectrum of the individual slices is then weighted according to the atom distribution. In our experiment, the averaged transition probability $\bar{W}^{02}(\nu)$ manifests itself in the spectrum derived from the cloud size after time of flight $\sigma(\nu) \propto \bar{W}^{02}(\nu)$.

The value extracted for the lattice depth from such a profile depends on the particular fit model. We use numerical simulations of $\bar{W}^{02}(\nu)$ with different parameters to test various fitting functions. The best agreement of the extracted lattice depth with the simulation input is achieved with
\begin{equation}
    \sigma(\nu) = \sigma_0 + \left(\frac{1}{2} + \frac{1}{\pi} \arctan\left(\frac{\nu-\nu_0}{\delta\nu}\right)\right)\left[k(\nu-\nu_0)+A\right], 
\end{equation}
where $\nu_0$ marks the position where the cloud width increases by half of the amplitude $A$. $\delta\nu$ sets the width of the step, $k$ sets the slope of the linear part above the step, and $\sigma_0$ is the cloud width below the band edge.

As a second method, we perform least-squares regression of the full numerical simulation of the experimental profiles. For this we vary $w_0$, $\sigma_c$, $r_c$, $V_0$, and $a$ such that the sum of the squares
\begin{equation}
    \sum_{\nu_i} \left[\sigma(\nu_i) - a \bar{W}^{02}(\nu_i)\right]^2
\end{equation}
across all measurement points $\nu_i$ is minimal. The effective lattice depth is then extracted by integrating $V(r)$ over the extent of the cloud. The results of the two methods usually agree within less than $4\%$.

\section{Systematic uncertainties from angle determination} \label{sec:appendix_B}
To address the issue of the impact of uncertainties in the angle determination, we rewrite Eqs.~(\ref{eq:STpol}) and (\ref{eq:meanpol}) as
\begin{equation}
    \tilde{\alpha}(\theta) = \tilde{\alpha}_0 + \frac{3}{4} \tilde{\alpha}_t\cos(2\theta).
\end{equation}
When measuring the polarizabilities $\tilde{\alpha}_\parallel$ and $\tilde{\alpha}_\bot$ for $\theta = 0$ and $\theta = \pi/2$, respectively, angle deviations of $\delta_\parallel$ and $\delta_\bot$ will result in measured values with systematic offsets corresponding to
\begin{equation}
    \begin{aligned}
    & \tilde{\alpha}'_\parallel = \tilde{\alpha}_0 + \frac{3}{4}\tilde{\alpha}_t\cos(2\delta_\parallel)\approx\tilde{\alpha}_0+\frac{3}{4}\tilde{\alpha}_t(1-2\delta_\parallel^2), \\
    &\tilde{\alpha}'_\bot = \tilde{\alpha}_0 + \frac{3}{4}\tilde{\alpha}_t\cos(\pi + 2\delta_\bot)\approx\tilde{\alpha}_0-\frac{3}{4}\tilde{\alpha}_t(1-2\delta_\bot^2).
    \end{aligned}
\end{equation}
When calculating the effect on the mean polarizability
\begin{equation}
    \tilde{\alpha}'_0 = \frac{1}{2}(\tilde{\alpha}'_\parallel+\tilde{\alpha}'_\bot) = \tilde{\alpha}_0 + \frac{3}{4}\tilde{\alpha}_t(\delta^2_\bot-\delta^2_\parallel),
\end{equation}
it becomes apparent that the errors will (partially) cancel each other. In particular, a systematic shift compared to the actual angles given by the magnetic field, such that $\delta_\bot=\delta_\parallel$, will cancel out completely.
In contrast, 
\begin{equation}
    \tilde{\alpha}'_t = \frac{2}{3}(\tilde{\alpha}'_\parallel-\tilde{\alpha}'_\bot) = \tilde{\alpha}_t(1- \delta^2_\bot-\delta^2_\parallel)
\end{equation}
is more sensitive to errors in the determination of $\theta$.





\begin{thebibliography}{57}
\bibitem [{Tec()}]{TechnicalReasonsNote}%
  \BibitemOpen
  \href@noop {} {}\bibinfo {note} {The stability of the lattice intensity, quality
  of the absorption imaging, and general stability of the experiment contribute
  to noise on the measurement. Furthermore, for technical reasons, the
  lattice depth is modulated already when ramping up the lattice.}\BibitemShut
  {Stop}%
\end{thebibliography}

%

\end{document}